%% file: Template.tex
\documentclass{article}
\usepackage{spconf,amsmath,graphicx}

\usepackage{graphbox}
\usepackage{caption,subcaption}
\usepackage{amssymb}
\usepackage{enumitem}
\usepackage{slantsc}
\usepackage{lipsum}
\usepackage{multirow}
\usepackage{dsfont}
\usepackage{color}
\usepackage{arydshln}
\usepackage{hyperref}
\usepackage{booktabs}
\usepackage{cite}
\usepackage[tracking=true]{microtype}

\title{CoRN: Co-trained Full- and No-Reference Speech Quality Assessment}

\name{Pranay Manocha$^{1}$,
      Donald Williamson$^{2}$,
      Adam Finkelstein$^{1}$}
\address{
  $^1$Department of Computer Science, Princeton University, Princeton, NJ, USA\\
  $^2$Department of Computer Science and Engineering, The Ohio State University, Columbus, OH, USA}
\DeclareUnicodeCharacter{2212}{-}
\input{macros}

\begin{document}
\ninept
\maketitle
\begin{abstract}
\input{texts/00-abstract}
\end{abstract}
\section{Introduction}
\label{sec:intro}
\input{texts/01-intro}

\section{Related Work}
\label{sec:background}
\input{texts/02-related}

\section{The CORN Framework}
\label{sec:format}
\input{texts/03-framework}

\section{Experimental Setup}
\label{sec:pagestyle}
\input{texts/04-experimental}

\vspace{-0.1in}
\section{Results}
\label{sec:typestyle}
\input{texts/05-results}

% \vspace{-0.1in}
\section{Conclusions and future work}
% \vspace{-0.1in}
\label{sec:majhead}
\input{texts/06-conclusion}

% \section{REFERENCES}
% \label{sec:refs}

\bibliographystyle{IEEEbib}
\bibliography{strings,refs}

\end{document}

%% file: macros.tex
\newcommand{\ignorethis } [1] {}

\newcommand{\eqnnum     } [1] {\mbox{(\ref{#1})}}

\newcommand{\eqn        } [1] {equation~\eqnnum{#1}}

\newcommand{\etal       }     {{et~al.}}

\newcommand{\eg         }     {{e.g.}}

\newcommand{\Reals      }     {{\textrm{I\kern-0.18em R}}}

\newcommand{\change     } [1] {\mbox{{\footnotesize $\Delta$} \kern-3pt}#1}

\newcommand\narrowstyle{\SetTracking{encoding=*}{-50}\lsstyle}
\newcommand\normalstyle{\SetTracking{encoding=*}{0}\lsstyle}

\newcommand{\acronym}[1] {\narrowstyle{\scshape{#1}}\normalstyle}

\newcommand{\PESQ}   {\acronym{Pesq}}
\newcommand{\VISQOL} {\acronym{Visqol}}
\newcommand{\POLQA}  {\acronym{Polqa}}

\newcommand{\SESQA}  {\acronym{Sesqa}}
\newcommand{\DPAM}   {\acronym{Dpam}}

\newcommand{\CDPAM}  {\acronym{Cdpam}}

\newcommand{\DNSMOS}  {\acronym{Dnsmos}}

\newcommand{\SNR}  {\acronym{Snr}}
\newcommand{\SISDR}  {\acronym{Si}-\acronym{sdr}}

\newcommand{\frameworkname}  {\acronym{Corn}}

\makeatletter
\newcommand*{\etc}{%
    \@ifnextchar{.}%
        {etc}%
        {etc.\@\xspace}%
}
\makeatother
\newcommand{\MOS}   {\acronym{Mos}}

\newcommand*{\skippingparagraph}{\par\vspace{0.5\baselineskip}\noindent}

%% file: texts/00-abstract.tex
Perceptual evaluation constitutes a crucial aspect of various audio-processing tasks. 
Full reference (FR) or similarity-based metrics rely on high-quality reference recordings, to which lower-quality or corrupted versions of the recording may be compared for evaluation.
In contrast, no-reference (NR) metrics evaluate a recording without relying on a reference.
Both the FR and NR approaches exhibit advantages and drawbacks relative to each other.
In this paper, we present a novel framework called \frameworkname\ that amalgamates these dual approaches, concurrently training both FR and NR models together. 
After training, the models can be applied independently.
We evaluate \frameworkname\ by predicting several common objective metrics and across two different architectures.
The NR model trained using \frameworkname\ has access to a reference recording during training, and thus, as one would expect, it consistently outperforms baseline NR models trained independently.
Perhaps even more remarkable is that the \frameworkname\ FR model also outperforms its baseline counterpart, even though it relies on the same training data and the same model architecture. 
Thus, a single training regime produces two independently useful models, each outperforming independently trained models.

% Our empirical findings suggest that the proposed framework surpasses autonomously trained models, underscoring the merits of fusing the strengths of these dual modalities. 
% %
% Thus, we posit that co-training a model alongside a more demanding constraint augments its efficacy within constraint-free scenarios.
%
%This framework holds the potential to significantly advance the development of audio processing metrics significantly, offering enhanced accuracy and applicability.

%% file: texts/01-intro.tex
% para about audio quality, and why is it such an important problem
% Quality assessment of speech signals plays a critical role in many applications. The gold standard for assessment of speech quality is subjective judgments by humans. Often, these subjective judgments are made by conducting different listening tests. Mean Opinion Score (MOS)~\cite{streijl2016mean} is the \emph{de facto} metric to assess speech quality through listening tests. However, such subjective evaluations are time and resource-consuming, especially when repeated many times per recording, and are therefore not scalable. Moreover, to obtain MOS reliably, one needs to control listening environments and hardware appropriately, further adding to the constraints of conducting MOS tests. This has led to considerable effort in developing alternatives to MOS tests.

Audio quality assessment plays a significant role across a variety of applications. Human judgment indicating how good or bad a clip sounds serves as the ``gold standard'' method for such evaluations. However, obtaining these judgments is resource-intensive due to the associated time and cost factors. Mean Opinion Score (\MOS)~\cite{streijl2016mean}, a widely used technique to gauge sound quality, demands substantial resources, especially when repeated many times per recording, and is therefore not scalable. Additionally, ensuring controlled listening conditions further compounds the challenges of conducting \MOS\ evaluations. Consequently, there exists a compelling impetus to explore alternative methodologies for quantifying sound quality.

% problem with existing metrics (full-reference)
Full-reference metrics, also known as intrusive or similarity metrics (e.g., \PESQ~\cite{rix2001perceptual}, \POLQA~\cite{beerends2013perceptual}, \VISQOL~\cite{hines2015visqol}, \DPAM~\cite{manocha2020differentiable}, \CDPAM~\cite{manocha2021cdpam}), require a clean reference to which a corrupted signal can be compared as the basis for a quality rating. Researchers commonly rely on full-reference metrics as a proxy for audio quality, because they were introduced earlier – consider, \eg\ SNR. One of the most impactful is \PESQ~\cite{rix2001perceptual}, introduced decades ago for telephony and still used today across tasks. 
These methods have been shown to correlate well with human perceptual judgments across tasks~\cite{manocha2020differentiable, manocha2021cdpam}.
%
%It correlates well with subjective listening tests~\cite{reddy2020dnsmos,fu2018end}, and \PESQ\ labels have also been leveraged for training \emph{differentiable} quality metrics~\cite{fu2018quality,fu2019metricgan,xu2021deep}.
%
However, a recent study by Manocha \etal~\cite{manocha2022audio} highlighted several real-world situations where established full-reference metrics (also known as similarity metrics) face discrepancies when compared with human perception. Specifically, their findings underscore that these metrics are unable to effectively account for the diverse range of audio quality variations in relation to ``clean" recordings created under distinct environments. Additionally, these metrics tend to accentuate differences that are virtually imperceptible. This phenomenon arises due to the metrics' training on pairs of recordings with identical speech content, resulting in models that lack robustness in distinguishing between alterations in content and variations in quality.

% problems with no-reference approaches
To ameliorate the reliance on clean reference, \emph{No-reference} methods rate quality on an absolute scale.
Traditional methods like ITU standard P.563~\cite{malfait2006p} and SRMRnorm~\cite{santos2014improved} involve complicated hand-crafted features.
State of the art approaches rely on deep  learning~\cite{reddy2020dnsmos,fu2018quality,gamper2019intrusive,yu2021metricnet,zhang2021end,mittag2021nisqa}.
%Based on neural networks, they are differentiable, making them suitable for 
%downstream tasks that also involve deep learning.
%
Earlier learning methods trained models on objective scores (\eg\ \PESQ)~\cite{fu2018quality},
while more recent approaches discover a mapping between noisy audio signals and \MOS\ in a supervised learning fashion~\cite{reddy2020dnsmos,dong2020pyramid,zhang2021end,catellier2020wawenets,mittag2021nisqa}.
However, as observed by Manocha \etal~\cite{manocha2021noresqa}, no-reference metrics learn an implicit distribution of clean references, which suffers from both (a) high variance due to factors like mood and past experience; and (b) substantial label variance from human annotations.
For example, in \DNSMOS~\cite{reddy2020dnsmos}, almost half of the recordings have ratings with standard deviation~$>1$. Such label noise poses challenges in training robust models. 
Given the pronounced variance within the training labels, the task of training robust models is further complicated, leading to instability and difficulties in achieving robust performance. Moreover, the progress in refining no-reference models consistently lags behind the advancements observed in full-reference model development, contributing to the intricate landscape of audio quality assessment.
%
% Collecting a consistent MOS dataset faces many challenges including the need for uniform listening setups across many subjects; and even then ratings may not be consistent when repeated. 
% %
% Finally, MOS ratings depend on the conditions of the stimuli, which are task-dependent
% (making it difficult to combine MOS ratings across tasks).
% multi-task learning a way to solve

This paper proposes to learn a model of speech
quality that combines multiple tasks. We call it \frameworkname\ for \emph{\underline{Co}-trained Full-\underline{R}eference and \underline{N}o-reference audio metrics}. \frameworkname\ learns from different types of tasks (FR and NR), 
%together with
%limitless amounts of 
%unlabeled or programmatically generated data, 
and produces speech
quality scores, together with usable latent features and informative auxiliary outputs. Scores and outputs are concurrently optimized in a multi-task setting by all the different speech quality assessment tasks, with the idea that each type of model outputs the same score irrespective of its handicap (with or without reference). 

As expected, the \frameworkname\ NR model demonstrates superior performance compared to an independently trained NR model. This advantage arises from co-training, which provides access to the reference during training and ensures stable training with consistent gradients. However, more remarkable is that the \frameworkname\ FR model surpasses its independently trained counterpart despite having the same architecture and training data. This outcome suggests that incorporating the NR loss during training assists the FR model in preventing over-generalization from the observed training content, thus enhancing its content-invariance.
By flowing information through a shared latent space bottleneck, the considered objectives learn to cooperate and promote better and more robust representations while discarding non-essential information (especially speech content information)~\cite{pascual2019learning}.
%The inspiration for the approach comes from human's ability to do the same. Given two completely random speech recordings, it is highly likely that a human would be able to compare them with respect to quality, irrespective of the actual speech content. 
%
%The hypothesis here is that adding an NR objective with the FR approach helps it learn content invariance, whereas adding an FR objective to the NR approach offers stable training and gradients. 

The \emph{key contributions} of this paper are: (1)~we propose a novel framework for speech quality assessment that produces \emph{both} FR and NR models, each capable of assessing sound quality independently of the other; (2)~we propose methods to train neural networks within this framework that are capable of predicting \SISDR~\cite{le2019sdr}, \SNR\cite{yuan2019signal} and \PESQ\cite{rix2001perceptual} scores both with and without reference recordings; and (3)~we evaluate our framework through several objective evaluations and show that the FR and NR models trained via \frameworkname\ outperform identical networks trained independently.

% \Adam{Maybe fold this in somewhere above... Benefits of CoRN approach include: 
% %
% (1) A single training regimen produces \emph{both} FR and NR models, each capable of assessing sound quality independently of the other. 
% %
% (2) Unsurprisingly, the CoRN NR model outperforms an equivalent NR model trained independently because during co-training affords access to the reference during training (maybe mention \emph{stability and gradients} as mentioned above?).
% %
% (3) However, it is more surprising that the CoRN FR model slightly outperforms an equivalent FR model trained independently; after all, they are trained with identical architecture and training data. It appears that the addition of the NR loss during training helps the FR model avoid over-generalizing from the content that it sees during training.}

\input{figures/main-diagram}

% We propose our metric, that combines both tasks together - and hence improves performance on both tasks together (as compared to trained independently) - say we do multi-objective learning, rather than multi-task learning

% self supervised, infinite data.

%% file: figures/main-diagram.tex
\begin{figure}[t!]
\vspace{-1\baselineskip}
\centering
\setlength{\tabcolsep}{4pt}
\includegraphics[width=\columnwidth]{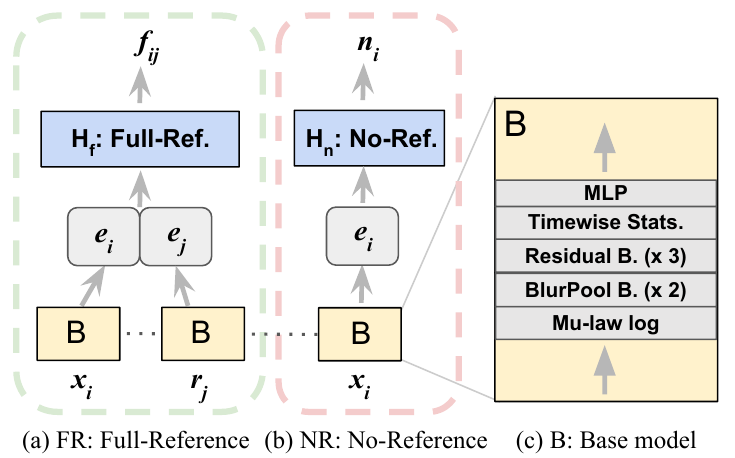}
%\vspace{-1.5\baselineskip}
\caption{Proposed CORN training framework with (a) Full-reference (FR, in green) and (b) No-Reference (NR, in red) models. Co-training (a) and (b) together -- the network architecture (c) of the base model $\mathbf{B}$ is identical in each instance in the FR and NR models, and has shared weights indicated by the dotted lines. In (a) and (b), task-specific output heads  $\mathbf{H}_f$ and $\mathbf{H}_n$ predict the FR and NR scores $f_{ij}$ and $n_i$. In FR the embedding $e_i$  of recording $x_i$ is identical to its counterpart in NR; however only in FR it is concatenated with the embedding $e_j$ of a reference recording $r_j$ before passing along to the output head (Sections~\ref{ssec: 3.1} and~\ref{subsec3.2}).}
\vspace{-1\baselineskip}
\label{model_architecture}
\vspace{-1ex}
\end{figure}

%% file: texts/02-related.tex
% full-reference model
\subsection{Full-reference metrics}
% \PESQ\cite{rix2001perceptual} and \VISQOL\cite{hines2015visqol} were some of the earliest models to approximate the human perception of audio quality.~Although useful, these had certain drawbacks: (i)~sensitivity to perceptually invariant transformations~\cite{hines2013robustness}; (ii)~narrow focus (\eg~telephony); and (iii)~non-differentiable, making it impossible to be optimized for as an objective in deep learning.~To overcome the last concern, researchers train differentiable models to approximate \PESQ~\cite{fu2019metricgan,zhang2018training}. The approach of Fu~\etal~\cite{fu2019metricgan} uses GANs to model \PESQ, whereas the approach of Zhang~\etal~\cite{zhang2018training} uses gradient approximations of \PESQ\ for training. Unfortunately, these approaches are not always optimizable, and may also fail to generalize to unseen perturbations.

% Instead of using conventional metrics (\eg~\PESQ) as a proxy, Manocha~\etal~\cite{manocha2020differentiable} proposed \DPAM\ that was trained directly on a new dataset of human JND judgments. \DPAM\ correlates well with human judgment for small perturbations, but requires a large set of annotated judgments to generalize well across unseen perturbations.
% %
% Most similar to our work is that of Serra \etal~\etal~\cite{serra2020sesqa} called \SESQA\ that was trained using the same JND dataset~\cite{manocha2020differentiable}, along with other objectives like \PESQ. However, it is yet to be investigated what happens when different types of tasks are added.

Early models (\PESQ\cite{rix2001perceptual}, \VISQOL\cite{hines2015visqol}) mimicked human audio quality perception, but had drawbacks: sensitivity to changes, narrow focus (e.g., telephony), and non-differentiability for deep learning. Researchers then trained differentiable models to imitate \PESQ\cite{fu2019metricgan}, using GANs or gradients. Yet, these methods had optimization and generalization issues.
Instead of using conventional metrics (e.g. \PESQ) as a proxy,
Manocha \etal\cite{manocha2020differentiable} proposed \DPAM\ that was trained directly on a
new dataset of human just-noticeable difference (JND) judgments. \DPAM\ correlates well with
human judgment for small perturbations, but requires a large set of
annotated labels to generalize well across unseen perturbations.
% An alternative emerged: Manocha \etal's \DPAM\cite{manocha2020differentiable}, trained directly on human Just Noticeable Differences (JND). While effective for small perturbations, \DPAM\ required extensive annotated data.
%
Similarly, Serra \etal's \SESQA\cite{serra2020sesqa} employed the JND dataset\cite{manocha2020differentiable}, adding objectives like \PESQ. However, the effects of adding different types of tasks remain unexplored.

\subsection{No-reference metrics}

% no-reference approach
Some of the earliest non-intrusive methods were based on complex hand-crafted, rule-based systems~\cite{malfait2006p,narwaria2011nonintrusive,sharma2014non}. 
Although they are automatic and interpretable, they tend to be task-specific, and do not generalize well. Moreover, these methods are non-differentiable which limits their uses within deep learning frameworks. To overcome the last concern, various neural network-based methods have been developed~\cite{reddy2020dnsmos,fu2018quality,andersen2018nonintrusive,lo2019mosnet,gamper2019intrusive,yu2021metricnet}. However, the issue of task-specificity and generalization remains. To overcome this, researchers proposed to train models directly on a dataset of human judgment scores~\cite{reddy2020dnsmos,lo2019mosnet,patton2016automos,dong2020pyramid}. Reddy \etal~\cite{reddy2020dnsmos} used a multi-stage self-teaching model~\cite{kumar2020sequential} to learn quality in the presence of noisy ratings. Nonetheless, no-reference metrics lag behind full-reference metrics in terms of correlation to human listening evaluations and adoption in practical cases.

% multi-task approach

%% file: texts/03-framework.tex
% framework
% \section{The Multi-Objective Framework}

Our framework,~\frameworkname\,, is designed to assess the quality of a given speech recording ($\mathbf{x}_\mathtt{i}$), an optional reference recording ($\mathbf{r}_\mathtt{j}$), and output a measure of FR $f_{ij}$ and NR $n_{i}$ quality. We propose a deep neural network for~\frameworkname\, and represent it by the function $\text{\frameworkname} = \mathcal{N}(x_{i},r_{j})$. Given that we do not rely on any human-labeled data, the crucial components of the framework include designing tasks and objective functions that can help learn a quality score.
Fig~\ref{model_architecture} is a simple illustration of the framework. 
In our current approach,~\frameworkname\, has the property of being monotonic (by design): if $\mathcal{N}(x_{a}, r_{j}) \geq \mathcal{N}(x_{b}, r_{j})$, 
%\Donald{This notation for xj1 and xj2 needs to be updated, since two subscripts are no longer used.} 
then $\mathtt{m}(x_{a}) \leq \mathtt{m}(x_{b})$, where $\mathtt{m}$ is any quality assessment measure as defined in Section~\ref{objective-metrics}. We do not enforce other metric properties~\cite{li2004similarity,chen2009similarity} to allow flexibility in defining tasks and objectives for training the neural networks. Moreover, even human judgment of similarity may not constitute a metric~\cite{tversky1977features}, and hence there is no pertinent reason which necessitates~\frameworkname\, to have metric properties. 
 
% architecture
\subsection{Framework Design and Model Architectures}
\label{ssec: 3.1}
\frameworkname\ architecture (Fig~\ref{model_architecture}) comprises two modules: a \emph{base model} block $\mathbf{B}$, and task specific \emph{output heads} $\mathbf{H}_f$ and $\mathbf{H}_n$. 

\skippingparagraph {\bf Base model block $\mathbf{B}$:}
We adopt a pre-existing model architecture inspired by \SESQA~\cite{serra2021sesqa}, and illustrated in Figure~\ref{model_architecture}(c). The model comprises of four primary stages. Initially, we pass the input $x$ through a $\mu$-law operation (without quantization) using a trainable $\mu$ parameter, which is initialized to 4. Following this, we utilize two pooling blocks, each comprising of convolution, batch normalization (BN), rectified linear unit (ReLU) activation, and BlurPool. These blocks employ 128 and 256 filters with a kernel width of 4, downsampling the input by a factor of 4. Subsequently, we employ three residual blocks. Each block consists of a BN preactivation followed by three stages of ReLU, convolution, and BN. These stages employ 512, 512, and 256 filters with kernel widths of 1, 3, and 1, respectively. A parametric linear averaging technique is employed to create a residual connection, expressed as follows: $h_0 = a_0h + (1 − a_0)F(h)$, where $a_0$ is a vector of adjustable parameters bounded within the range of 0 to 1, $F$ denotes the residual network, $h$ signifies the input to the residual layers, and $a$ represents the weight associated with the output and the residual layer. The components of $a$ are initialized to 6, ensuring an initial emphasis on a direct path from $h$ to $h_0$. 
%\Donald{What are a and h? How are they generated?}
After the residual blocks, temporal statistics are computed on a per-time basis, involving channel-wise mean and standard deviation calculations. This process consolidates all temporal information into a singular vector of dimensions 2×256. The vector undergoes BN before being fed into a multi-layer perceptron (MLP) comprising two linear layers with BN, and a ReLU activation in between. This MLP consists of 1024 and 200 units.
We further show that our results/hypothesis do not change if the architecture changes (refer to Sec~\ref{invariance-base-model}).
The next blocks consist of output heads for the training tasks and are described below, along with the training loss functions. 

% Multi-objective learning
\subsection{Training Tasks and Loss Functions}
\label{subsec3.2}

\noindent {\bf FR Block:}
As shown in Figure~\ref{model_architecture}(a), this block
is designed such that the base network takes the two inputs ($x_{i}$ and $r_{j}$),  concatenates their embeddings ($e_i$ and $e_j$) and feeds it further onto two shallow linear layers $H_f$ that predict $f_{ij}$, the Scale-Invariant Signal to Distortion Ratio (\SISDR) for the entire recording. 

The goal of this task is to predict \SISDR. 
%\Donald{I'm starting to think that we should just use s to represent the objective label (e.g., instead of shat) and shat to represent the network output, instead of r. r is used for the reference above and can be confusing, and the hat is generally used to represent an estimate. Additionally, it's worth noting that the label for the NR and FR blocks are the same, since the notation may lend one to believe that they differ. }
Let $\hat{s}=f_{ij}$  be the recording level \SISDR\ predicted by this output head. We then use the \textit{Smoothed-L1} loss between $\hat{s}=f_{ij}$ and the target \SISDR\ $s$ to train the network:

% \Donald{Since space is an issue, it may help to consolidate equations (1) and (2) into a single equation, since all that changes is $nr_i$ and $fr_i$.}
\begin{equation}
\label{eq-1}
L_Q(s, \hat{s}) = \begin{cases}
(\hat{s} - s)^{2}/\beta, & |\hat{s} - s| \leq \beta\\
2|\hat{s} - s| - \beta, & \text{otherwise}
\end{cases}
\end{equation}

\skippingparagraph {\bf NR Block:}
As shown in Figure~\ref{model_architecture}(b), this block is designed such that the base network takes a single input $x_{i}$, passes it through the same base model B to produce the same embedding $e_i$ as found by the FR path.
Next it feeds further onto two shallow linear layers $H_n$ that predict $\mathbf{n}_{i}$, the SI-SDR for the entire recording. 
%The goal of this task is to predict the SI-SDR. 
Using $\hat{s}=n_{i}$ (predicted SI-SDR) we apply the \textit{Smoothed-L1} loss between predicted and target SI-SDR $s$ to train the network, as in \eqn{eq-1}.

% \begin{equation}
% L_Q(\mathbf{nr_{i}}, \mathbf{s}_{i}) = \begin{cases}
% (nr_{i} - s_{i})^{2}/\beta, & \|nr_{i} - s_{i}\| \leq \beta\\
% 2\|nr_{i} - s_{i}\| - \beta, & \text{otherwise}
% \end{cases}
% \end{equation}

\skippingparagraph {\bf Objective Metrics:}
\label{objective-metrics}
Since we do not have any perceptual labels, ~\frameworkname\, relies on a signal processing measure \SISDR, to compare the quality of the two inputs. 
We consider \SISDR\ objective metric as a proxy for human quality evaluation because it helps us train on limitless amounts of unlabeled or programmatically generated data and outputs quality scores that are consistent, unlike \MOS.
% In this paper, we propose to learn a model of speech
% quality that combines multiple tasks, following a self-supervised approach. We call it \frameworkname\ for \emph{\underline{Co}-trained Full-\underline{R}eference and \underline{N}o-reference audio metrics}. \frameworkname\ learns from different types of tasks (FR and NR), together with
% %limitless amounts of 
% unlabeled or programmatically generated data, and produces speech
% quality scores, together with usable latent features and informative auxiliary outputs. Scores and outputs are concurrently optimized in a multi-task setting by all the different speech quality assessment tasks, with the idea that each type of model outputs the same score irrespective of its handicap (with or without reference). 
%
\SISDR~\cite{le2019sdr} is a measure that was introduced to evaluate performance of speech-processing algorithms. It is invariant to the scale of the processed signal and can be used to quantify quality in diverse cases, including additive background noises as well as other distortions. 
Additionally, we also show that our framework is invariant to the target label we use, so we show its performance on other objective metrics like \SNR\ and \PESQ\ (refer to Sec~\ref{invar-target-obj}).

\SNR\ is measured as the ratio of the signal power to the noise power and is primarily meant only for additive noises. 
Consider a mixture signal $x$, 
$x = r + \delta \in 
\mathbb{R}^L$ 
where $r$ is the clean signal and $\delta$ is the noise signal, then
\begin{equation}
\mbox{SNR} = 10\log_{10} \left(\frac{||r||^2}{||r - x||^2}\right)
\label{snr}
\end{equation}
$10log_{10} ()$ factor measures SNR in dB-scale, and a higher SNR implies better signal quality. Yuan et al.~\cite{yuan2019signal} also showed that SNR as a distance metric had better properties than conventional metrics (like Euclidean distance).

\PESQ~\cite{rix2001perceptual}, which stands for \emph{perceptual evaluation of speech quality,} is an impactful objective metric used by many researchers to evaluate the sound quality of their model output with respect to a given reference. It was introduced decades ago for telephony and still used today for a wide variety of tasks including enhancement~\cite{su2020hifi,defossez2020real,zezario2021deep,Su:2021:HSS}, vocoders~\cite{cernak2005evaluation}, and transmission codecs~\cite{beerends2004measurement,paglierani2007uncertainty}.

% Training procedure
\subsection{Training procedure}
\label{training_procedure}
We now describe our training procedure. We assume the availability of a clean speech database $\mathcal{D}_\text{clean}$. The training inputs $\mathbf{x}_{i}$ and $\mathbf{r}_{j}$ are created by sampling a clean recording $r_j$ from $\mathcal{D}_\text{clean}$. $r_{j}$ is corrupted to produce $x_{i}$. The degradations we use can be largely grouped under two categories~(a) additive noise degradations, and~(b) speech distortions based on signal manipulations - \emph{Clipping}, \emph{Frequency Masking}, and \emph{Mu-law compression}. For additive noise, we sample noise recordings, $\delta_i$, from a noise database (Section~\ref{ssec: datasets_and_training}) and add them to $r_j$ at SI-SDR levels uniformly sampled from the range -40 dB to +40 dB. Once we have the degraded signal ($x_i$), its clean-reference counterpart ($r_j$), and their quality score $\mathbf{f}_{ij}$ and $\mathbf{n}_{i}$(=$\mathbf{f}_{ij}$), we can train the network as described in previous sections. 

%Note that, for degradations under the second category, the \emph{quantification-task} loss $L_Q$, includes loss only from the SI-SDR head, $L_{sdr}$. SNR cannot be accurately computed for these distortions (residual $x-s$ may not be orthogonal to $s$) and hence we rely only on SI-SDR for training the network in these cases.

% Usage
\subsection{Inference}
\label{usage}

Once the network is trained, we can predict the quality score of a test input $x_{\text{i}}$ along with the option of an accompanying reference input, $r_{\text{j}}$. Within this framework, our Full-Reference (FR) branch accepts two inputs, namely, $x_{\text{i}}$ and $r_{\text{j}}$, and generates an output quality score, $f_{\text{ij}}$. Likewise, our No-Reference (NR) branch is designed to process a solitary input, $x_{\text{i}}$, and produces an NR quality score, $n_{\text{i}}$. The selection between these branches hinges on whether our task incorporates a reference, allowing us to determine the appropriate branch for utilization.

\input{tables/main-table}

%% file: tables/main-table.tex
\begin{table}[b!]

\vspace{-0ex}
\centering
\renewcommand{\arraystretch}{1.1}
\resizebox{\columnwidth}{!}{
 \begin{tabular}{l c c c c c c : c c}
 \toprule
    \multirow{2}{*}{\bf Name} & \multicolumn{2}{c}{\bf SI-SDR} &  \multicolumn{2}{c}{\bf SNR} & \multicolumn{2}{c}{\bf PESQ} & \multicolumn{2}{c}{\bf New $\mathbf{B}$ Arch.}\\
 \cmidrule(lr){2-3} \cmidrule(lr){4-5} \cmidrule(lr){6-7} \cmidrule(lr){8-9}
 & \bf FR &\bf NR & \bf FR &\bf NR & \bf FR &\bf NR & \bf FR &\bf NR\\
 \cmidrule(lr){1-9}

  {\bf Indiv.} & 96.5 & 110.3  & 98.0 & 99.9 & 0.9 & 1.3 & 124.9 & 134.9  \\
  {\bf \frameworkname} & \bf 85.9  & \bf 92.9 & \bf 79.5 & \bf 82.7 & \bf 0.7 & \bf 0.9 & \bf 103.2 & \bf 108.2 \\
    
 \bottomrule
\end{tabular}
}
\caption{\textbf{Evaluations}: Refer to Sec~\ref{ssec-held-out}. Models include: CORN and individual FR and NR prediction models. The numbers show SI-SDR as a metric, unless specified otherwise. $\downarrow$ is better.}
\label{tab:mos_1}
\end{table}

% Metric Performance
% SI-SDR, SNR and PESQ

% Si-SDR proxy at the right vertical dotted line

%% file: texts/04-experimental.tex
% \section{Experimental Setup}

\subsection{Datasets and training}
\label{ssec: datasets_and_training}
For training ($\mathcal{D}_{\text{train}}$), the clean set $\mathcal{D}_{\text{clean}}$ comes from the DAPS dataset~\cite{mysore2014can}, and the noise set $\mathcal{D}_{\text{noise}}$ comes from DNS Challenge ~\cite{reddy2020interspeech} dataset. Along with additive noise, clipping, and frequency masking distortions are used during training. For robustness and better generalization to realistic conditions, we also add reverberation using room impulse responses from the DNS Challenge dataset.
For the test-set ($\mathcal{D}_{\text{test}}$), we use TIMIT~\cite{garofolo1993darpa} as the source for clean speech, and ESC-50~\cite{piczak2015esc} dataset for noise recordings. The test set also includes Gaussian noise addition and Mu-law compression as unseen degradations.
The inputs to our model are 3-second waveform excerpts. We use the Adam optimizer with a learning rate of $10^{−4}$ with a batch size of 64. We train the network for 1000 epochs. Smoothed L1 parameter $\beta$=1 for all experiments. 
%
%\Donald{It's not clear what is meant by "Held-Out" and "Invar. to Base M." in Table 1. These may need to be rephrased. Furthermore, the way the results are presented gives the impression that the approach is not adequately compared with other approaches, because there are only two rows. It may be worth puting the FR and NR approaches on separate rows and/or separating the SI-SDR results from that of SNR/PESQ.}
\subsection{Baselines}
We undertake a comparative analysis involving our proposed methodology, \frameworkname, in contrast to standalone models that were trained without adopting a multi-task framework, encompassing individual Full-Reference (FR) and No-Reference (NR) models. This endeavor aims to ascertain the potential superiority of our amalgamated model, \frameworkname.

%% file: texts/05-results.tex
% subjective analysis: correlation with show performance with GT OOD datasets on SI-SDR predictions
% \input{tables/ablation}
\subsection{Performance across metrics and architectures}
\label{ssec-held-out}
% \input{tables/held-out}

% \Adam{round these numbers to the nearest integer, or at least no more than one digit after decimal (no more precision needed, and completely understandable at integer). Same for Tables 2 and 4. Tables 3,5,6 numbers fine.}

These aim to assess \frameworkname\ as a proxy for subjective judgments by humans. More specifically, we evaluate how well \frameworkname\ correlates with ground truth target objective measures. We first hold out a subset of $\mathcal{D}_\text{train}$ and evaluate the performance of the models on that set. Next, to show generalization to unseen conditions (room environments, listeners, etc.), we also evaluate models across the unseen test dataset $\mathcal{D}_\text{test}$. 
We evaluate various models based on the output from the model, compared to the ground truth noise level using mean square error (MSE) between noise level differences.

Results are displayed in Table~\ref{tab:mos_1}. The proposed  \frameworkname\ framework demonstrates superior performance compared to the individually trained baseline models. When using \frameworkname\ for training, the no-reference (NR) model shows an improvement of around 16\% over the independently trained NR model, suggesting that co-training stabilizes the NR model.
Perhaps more remarkably, the full-reference (FR) model trained with \frameworkname\ also exhibits a sizeable gain, surpassing the individually trained FR model by around 11\%, despite the fact that the independent and co-trained FR models share the \emph{exact} same architecture and training data.
%
%\Donald{It may help to be a little more specific here and clearly state the performance differences.} 
Thus, the co-training approach enhances the quality of each of the respective NR and FR models that were trained together.

\input{tables/ablation}

% \subsection{Generality of \frameworkname\,}
% \label{ssec-genrality}

Objectively, we also compare various metrics on (i) invariance to base model architecture; and (ii) invariance to target objective.

\skippingparagraph {\bf Invariance to base model}
\label{invariance-base-model}
To demonstrate the generality of our proposed framework, transcending reliance on any specific architectural paradigm, we effectuate a substitution of the base model $\mathcal{B}$ with an architecture introduced by Manocha \etal~\cite{manocha2021noresqa}. This alternative model employs a composite of both magnitude and phase spectra as input. Please refer to Table~\ref{tab:mos_1} for detailed insights. It becomes evident from the outcomes illustrated in Table~\ref{tab:mos_1} that the performance of this model aligns coherently with those derived from the model trained using the base model architecture described in Sec~\ref{ssec: 3.1}.

\skippingparagraph {\bf Invariance to choice of target objective}
\label{invar-target-obj}
In order to establish the universality of our proposed framework, devoid of any reliance on specific target objectives, we enact a substitution of the initial SI-SDR training objective with PESQ and SNR. For a fair comparison, given that SNR is confined to linear degradations, such as background noise, we adapt our approach to only introduce diverse background noise types during training and evaluation. For comprehensive insights, kindly refer to Table~\ref{tab:mos_1}. The observations from the metrics presented therein harmonize with outcomes originating from the SI-SDR paradigm, underscoring its efficacy across various objectives. It can be noted that conducting training using the \PESQ\ metric results in reduced error rates. This outcome may be attributed to the fact that \PESQ\ has a scale ranging from 1 to 5, whereas \SNR\ and \SISDR\ exhibit a broader range spanning from -40 dB to +40 dB.
% explain about SNR, and how it is different in terms of dataset training.

\subsection{Evaluation of the embedding}
\label{ssec-eval-embed}

\noindent {\bf Content invariance}
% \input{tables/content-invariance}
% We next evaluate the in-variance and robustness of \modelname\ to speaker’s gender. Moreover, for a given test recording, it does not matter whether the speaker’s gender in the reference is same or not. This supports our key hypothesis that non-matching references are sufficient for depth assessment.
% To evaluate the robustness to content variations, we create a test dataset of two groups: the first group consists of pairs of recordings with the same noise but different speech content; the other group consists of recordings with different noise and speech content. We use the output of the base model $e_i = \mathbf{B (x_i)}$ as embeddings and then compute the cosine similarity between the pair of recordings for both groups, and then fit a gaussian distribution over the samples from the groups.
% %
% We calculate the common area between these normalized gaussian distributions. 
% %\Donald{How is the common area calculated? Is there a reference? How are the gaussian distributions computed? Are they computed from the embeddings?}
% %
% The smaller the common area, the more robust the model. Our proposed model has the lowest common area (across both FR and NR), suggesting in-variance to content (refer to Table~\ref{tab:mos_2}).
% %
% Decreasing common area also corresponds with better performance on the held-out dataset (Table~\ref{tab:mos_1}), suggesting that the task of separating these two distribution groups may be a factor when learning a robust audio quality assessment metric.
%
To assess the robustness of the system to content variations, we generate a test dataset encompassing two distinct groups: the first group comprises pairs of recordings characterized by identical background noise but different speech content, while the latter group comprises recordings featuring different noise and speech content. Embeddings $e$ are extracted from the base model $\mathbf{B}$, and subsequently, the cosine similarity between recording pairs from both groups is computed. A Gaussian distribution is then fitted to the resultant samples drawn from these groups.

The calculation of the common region between these normalized Gaussian distributions is performed to gauge the extent of their overlap. A smaller common area is indicative of a more robust model. In \frameworkname\, it is noteworthy that the common area is observed to be the lowest across both scenarios, namely with and without content variations (as delineated in Table~\ref{tab:mos_2}).

Moreover, it is worth highlighting that a diminishing common area is associated with enhanced performance on the held-out dataset, as presented in Table~\ref{tab:mos_1}. This finding suggests that the challenge of distinguishing between these two distribution groups may exert a significant influence on the process of acquiring a robust audio quality assessment metric.

% show for both FR and NR

% \noindent {\bf Commutative property}
% We empirically study how well our framework supports the commutative property - the estimate should stay the same for $\mathcal{N}(\xvect_A, \xvect_B)$ and $\mathcal{N}(\xvect_B, \xvect_A)$. Changing the order of inputs should not change the \frameworkname\ score. We find that for only a small fraction (less than 2.5$\%$) of the test pairs, changing the order also changes the \frameworkname\ score by more than 0.15m. Preference task outputs remain consistent as well (flips on changing the order) for more than 97$\%$ of the test pairs.

\skippingparagraph {\bf Small shifts in signal}
To evaluate the robustness to small (imperceptible) signal shifts, we create a test dataset of pairs of recordings with clean references and small noise-added signals. It is anticipated that the FR outputs shall closely approximate the highest attainable scores across a majority of the test instances. Conversely, in the context of NR scores, minimal disparities between the two input signals are sought, with the aim of these disparities approaching proximity to zero. Refer to Table~\ref{tab:mos_2}. Here, we show the difference between the maximum score and the model FR outputs for the FR case and the magnitude difference of the respective scores for the NR case. In all cases, we see that our model has the lowest scores, showing that the model is robust to small, imperceptible changes.

% show for both FR and NR

\skippingparagraph {\bf Quality based retrieval}:
Here, we consider the outputs after the base model block as the quality embeddings, and use it for quality-based retrievals. Similar to Manocha~\etal~\cite{manocha2021noresqa}, we first create a test dataset of 1000 recordings at 10 discrete quality levels. We take randomly selected queries and calculate the number of correct class instances in the top $K$ retrievals. We report the mean of this metric over all queries ($\text{MP}^k$). \frameworkname\ gets $\text{MP}^{k=10}$~=~0.87, as compared to $\text{MP}^{k=10}$~=~0.75 for the FR model, and $\text{MP}^{k=10}$~=~0.80 for the NR model. This suggests that our approach better clusters quality-level groups in this learned space.
% show for both FR and NR

%% file: tables/ablation.tex
\begin{table}[b!]

\vspace{-1\baselineskip}
\centering
\renewcommand{\arraystretch}{1.0}
\resizebox{0.8\columnwidth}{!}{
 \begin{tabular}{l c c c c}
 \toprule
    \multirow{2}{*}{\bf Name} & \multicolumn{2}{c}{\bf Invar. to content} & \multicolumn{2}{c}{\bf Small signal shifts}\\
 \cmidrule(lr){2-3} \cmidrule(lr){4-5}
 & \bf FR &\bf NR & \bf FR &\bf NR \\
 \cmidrule(lr){1-5}

  {\bf Indiv.} & 0.5 & 0.8 & 1.4 & 2.0    \\
  {\bf \frameworkname} & \bf 0.3 & \bf0.3 & \bf 1.2 & \bf 1.9 \\
    
 \bottomrule
\end{tabular}
}
\caption{\textbf{Evaluations}: Refer to Sec~\ref{ssec-eval-embed}. Models include: CORN and individual FR and NR prediction models. $\downarrow$ is better.}
\label{tab:mos_2}
\end{table}

%% file: texts/06-conclusion.tex
This paper presents \frameworkname\ -- a novel approach that co-trains FR and NR models. We find that incorporating the NR loss during training assists the FR model in preventing over-generalization from the observed training content, thus enhancing its content-invariance. On the other hand, incorporating the FR loss during training assists the NR model by providing stable gradients during training.  
%
% This approach also holds the potential to further advance the development of other audio quality metrics, by offering enhanced robustness and applicability. 

In the future, we would like to apply this framework to a broader set of objectives and quality metrics. For example,
we believe it would be valuable to collect a large dataset of human subjective ratings like \MOS\ in a format suitable for training FR and NR models with such data. Likewise, the framework could be extended to learn from non-scalar data such as pairwise preference or triplet judgments.

%\Donald{Based on the page requirements, it seems like approximately 20 references need to be removed, to get within the 5-page limit. Additionally, a space is needed in-between the author's initial and last name, just to help with readability.}